\documentclass[fleqn,12pt,twoside]{article}
\usepackage{espcrc1}
\input epsf
\title{Nucleon Structure in the Resonance Region \footnote{Talk presented 
at the International Conference on Mesons amd Light Nuclei, 
Prague, July 2 - 6, 2001}}

\author{Volker D. Burkert \\
Thomas Jefferson National Accelerator Facility\\
12000 Jefferson Avenue, Newport News, VA23606, USA\\
{\small email: burkert@jlab.org}}

\begin{document}
\maketitle

\abstract{I discuss recent results of inclusive and exclusive 
electroproduction experiments 
at Jefferson Lab. They include measurements of the spin response 
for protons and neutrons in the resonance region, 
exclusive single pion and multiple pion production to measure 
resonance transition
multipoles, and searches for missing quark model states. A brief 
outlook to the new domain of Generalized Parton Distributions is given
as well.}

\section{Introduction}

Studies of the nucleon structure for over 30 years
have focused
on the deep inelastic regime to determine the quark momentum and spin 
distributions, and to test fundamental sum rules. One of the surprising 
findings was that less than $25\%$ of the nucleon spin is accounted 
for by the spin of quarks \cite{filipone}. 
This result is in strong contradiction to expectations, which shows 
that we are far from having a realistic
picture of the intrinsic structure of the nucleon. 
Moreover, the nucleon structure has hardly been 
explored in the regime of confinement, which is the true domain
of strong QCD. Our understanding of nucleon structure is not complete 
if the nucleon is not also probed and fundamentally described 
at large or medium distances. 
This is the domain where current experiments at JLab have their 
biggest impact.           
It is only through a concerted effort of precise experiments and new 
approaches in theory that we will be able to understand nucleon structure
from the smallest to the largest distances within a consistent framework. 
Experiments at JLab aim at providing precise data 
as the basis for such an endeavor.

\begin{figure}
\vspace{65mm} 
\centering{\includegraphics{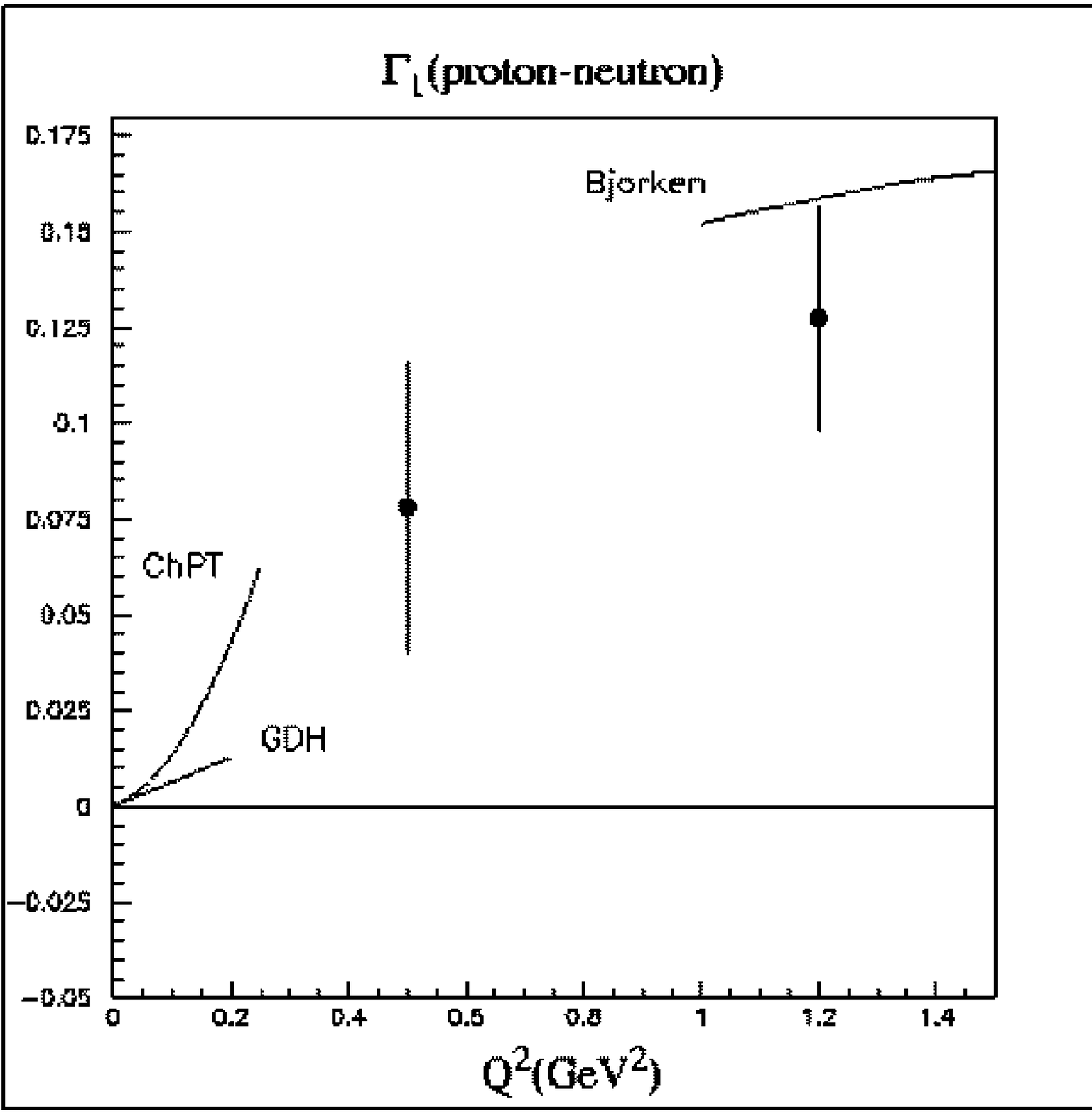}}
\centering{\includegraphics{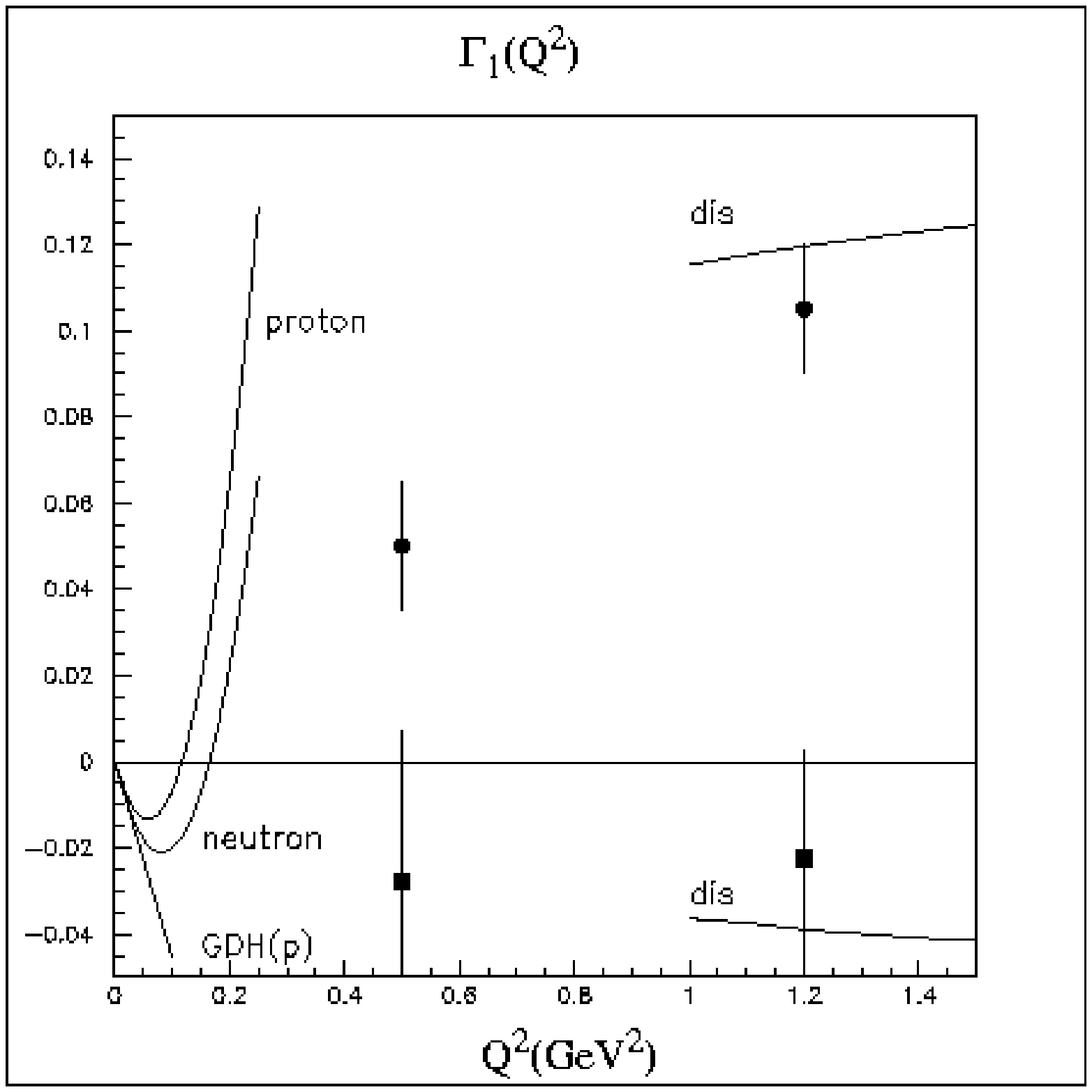}}
\caption{\small First moments of the spin structure 
function $g_1(x,Q^2)$ for the 
proton and neutron (left), and for the proton-neutron 
difference (right). The curves above $Q^2=1$GeV$^2$ are pQCD evolutions of 
the measured $\Gamma_1$ for proton and neutron, and the pQCD evolution for 
the Bjorken sum rule, respectively. The straight lines near $Q^2=0$ indicate 
the slopes given by the GDH sum rule. 
The curves at small $Q^2$ represent the NLO HBChPT results.}.
\label{gamma1}
\end{figure}

\section{Spin Response of the Proton and Neutron}

The inclusive doubly polarized electron-nucleon 
cross section can be written as:
$${1\over \Gamma_T} {d\sigma \over d\Omega dE^{\prime}} = \sigma_T 
+ \epsilon\sigma_L + P_eP_t[\sqrt{1-\epsilon^2}A_1\sigma_T\cos{\psi} + 
\sqrt{2\epsilon(1+\epsilon)}A_2\sigma_T\sin{\psi}]~\eqno(1)$$
where $A_1$ and $A_2$ are the spin-dependent asymmetries, $\psi$ 
is the angle between the nucleon polarization vector and the 
$\vec q$ vector, $\epsilon$ the
polarization parameter of the virtual photon, and $\sigma_T$ and $\sigma_L$ 
are the total absorption cross sections for transverse and longitudinal 
virtual photons. Experiments usually measure the asymmetry 
$$A_{exp} = P_eP_tD {A_1+\eta A_2 \over 1+\epsilon R}\eqno(2)$$
where D is a kinematical factor describing the polarization transfer from the
electron to the photon. 
$A_1$ and $A_2$ are related to the spin structure function $g_1$ 
by  $$g_1(x,Q^2) = {\tau \over 1+\tau}
[A_1 + {1\over \sqrt{\tau}}A_2]F_1(x,Q^2) \eqno(3)$$ 
where $F_1$ is the usual unpolarized structure function, and 
$\tau \equiv {\nu^2 \over Q^2}$. 

\begin{figure}
\vspace{90mm} 
\centering{\includegraphics{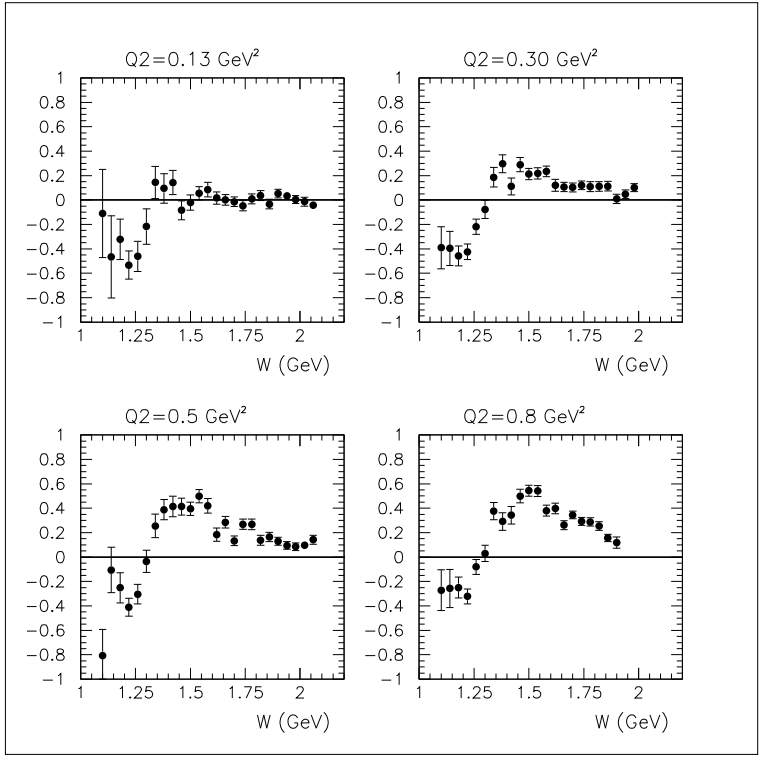}}
\caption{\small Asymmetry $A_1+\eta A_2$ for protons. The 
panels show preliminary results from CLAS at a beam energy of 2.6 GeV and
for different $Q^2$ values.}
\label{epasym}
\end{figure}
 
An important quantity is the first moment 
$\Gamma_1(Q^2) = \int{g_1(x,Q^2)}dx$.
The Gerasimov-Drell-Hearn (GDH) sum rule \cite{gerasimov,drell}, and Bjorken sum rule 
$\Gamma_1^p-\Gamma_1^n = { 1\over 6}g_A$ for the proton-neutron difference, 
provide constraints for $\Gamma_1$ at the kinematical 
endpoints $Q^2 \rightarrow 0$, and $Q^2 \rightarrow \infty$. 
The evolution of the Bjorken sum rule to finite values of $Q^2$ using 
pQCD and the Operator-Product-Expansion (OPE) connects 
experimental values measured at
finite $Q^2$ to the endpoint. At the opposite end, the GDH sum rule
defines the slope of $\Gamma_1$: 
$$2M^2{d\Gamma_1 \over dQ^2}(Q^2\rightarrow 0) = -{1\over 4}\kappa^2 \eqno(4)$$
where $\kappa$ is the anomalous magnetic moment of the target nucleon.  
Heavy Baryon Chiral Perturbation Theory (HBChPT) may be used 
to evolve the GDH sum rule to $Q^2 \neq 0$ \cite{xji}. 
The challenge of nucleon structure physics is to test the validity of these 
evolutions, and to bridge the remaining 
gap. Lattice QCD may play an important role in
describing resonance contributions to the moments of spin 
structure functions.  
Using just the constraints given by the two endpoint sum rules we may already
get a qualitative picture of $\Gamma^p_1(Q^2)$ and $\Gamma^n_1(Q^2)$. 
There is no sum rule for the proton and neutron separately that
has been verified, however, experiments have determined the asymptotic 
limits with sufficient confidence for the proton and the neutron. At large 
$Q^2$, $\Gamma_1$  is expected to approach this limit following the 
pQCD evolution from finite values of $Q^2$. At small $Q^2$, $\Gamma_1$ 
must approach zero with a slope given by the GDH sum rule 
(assuming the sum rule will be verified). 
The situation is depicted in Figure {\ref{gamma1}}, where also 
the next-to-leading HBChPT evolution at small $Q^2$ and the 
pQCD evolution to order $\alpha_s^3$ at high $Q^2$ are shown. 
\begin{figure}
\vspace{65mm} 
\centering{\includegraphics{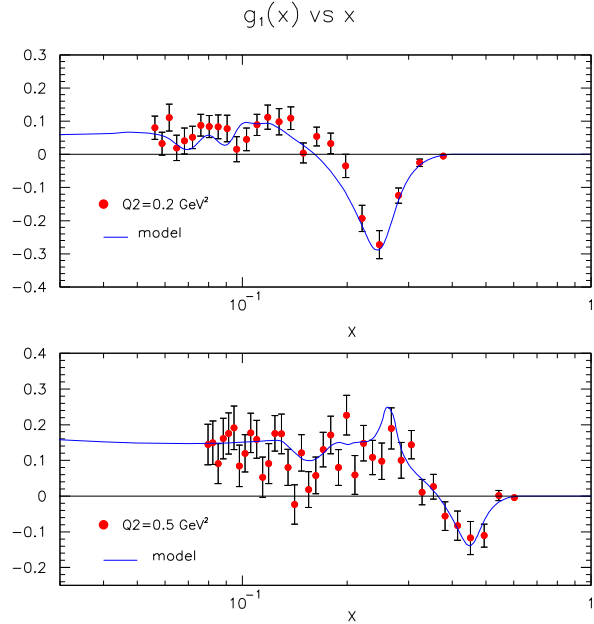}}
\caption{\small Preliminary CLAS results on the spin structure 
function $g_1(x,Q^2)$ for the 
proton. The curve labeled ``model'' is used for radiative corrections, and 
to extrapolate to $x=0$ for the evaluation of $\Gamma_1$.}
\label{g1x}
\end{figure}
As the slope at $Q^2=0$ is $< 0$, and the asymptotic value is $>0$, 
$\Gamma^p_1$ must change sign at 
some value $Q^2 < 1$ GeV$^2$. 
We note that the HBChPT evolution \cite{xji} 
cannot give a good description of the 
trend shown by the existing data, for $Q^2> 0.1$ GeV$^2$. 
However, for the proton-neutron difference the situation is quite 
different \cite{burk}; the HBChPT curve describes the 
general trend of the data quite well, and over a significantly 
larger range in $Q^2$ than for proton and neutron separately.

\subsection{The first moment $\Gamma_{1}(Q^2)$ for the proton.}

Inclusive double polarization experiments have been carried out on 
polarized hydrogen \cite{bucramin} 
using $N\vec {H_3}$ as polarized target material. 
In Figure {\ref{epasym}} the asymmetry is shown for various bins in $Q^2$. 
For the lowest $Q^2$ bin the asymmetry is dominated by 
the excitation of the $\Delta(1232)$, resulting in a strong negative asymmetry. 
At higher $Q^2$ the asymmetry in the
$\Delta(1232)$ region remains negative, but quickly becomes positive
and large at higher W, reaching peak values of about 
0.6 at $Q^2=0.8$ GeV$^2$ and W=1.5 GeV. Evaluations of resonance
contributions show that this is largely driven by the 
$S_{11}(1535)$ $A_{1/2}$ amplitude, and by the rapidly changing helicity 
structure of the strong $D_{13}(1520)$ state. The latter resonance is 
known to have a dominant $A_{3/2}$ amplitude at the photon point, but 
is rapidly changing to $A_{1/2}$ dominance for 
$Q^2 > 0.5$ GeV$^2$ \cite{burk1}.

\begin{figure}[t]
\vspace{90mm} 
\centering{\includegraphics{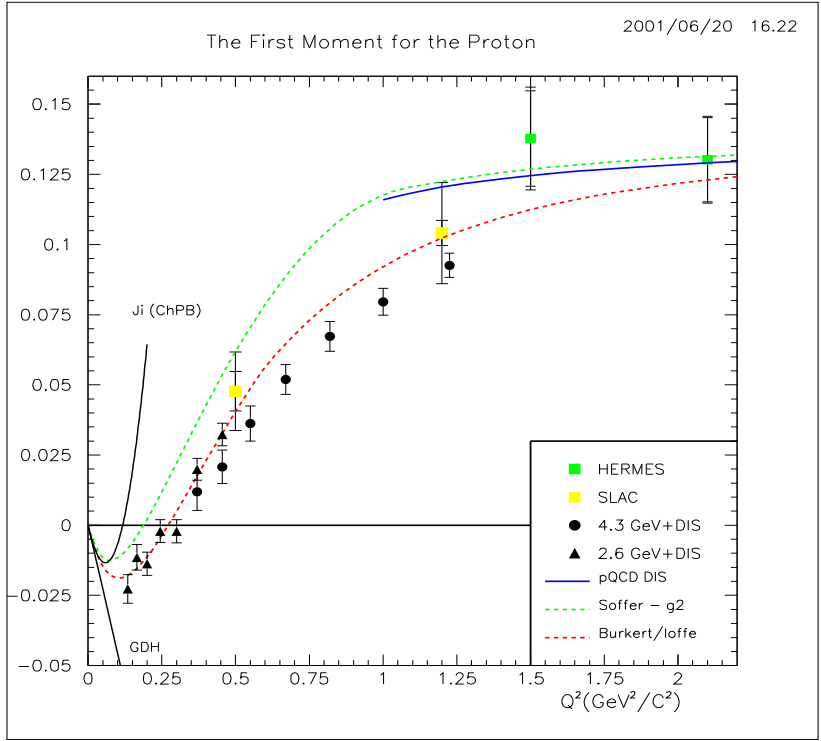}}
\caption{\small The first moment 
$\Gamma_1(Q^2)$ for the proton. The full symbols are preliminary 
results from CLAS. Data from SLAC are shown for 
comparison. The curves are from ref. \cite{soffer}, \cite{buriof}}
\label{gammafull}
\end{figure}

Using a parametrization of world data on $F_1(x,Q^2)$ and 
$A_2(x,Q^2)$ we can extract $g_1(x,Q^2)$ from (5). Examples of 
$g_1(x,Q^2)$ are shown in Figure {\ref{g1x}}. 
The main feature at low $Q^2$ is due to the negative contribution of the
$\Delta(1232)$ resonance. The graphs also show a model parametrization 
of $g_1(x,Q^2)$ which was used to extrapolate to $x \rightarrow 0$. 
The extrapolation is needed to evaluate the first moment $\Gamma_1(Q^2)$ 
which is shown in Figure {\ref{gammafull}}. The 
characteristic feature is the strong $Q^2$ dependence for $Q^2 < 1$ GeV$^2$, 
with a zero crossing near $Q^2=0.3$ GeV$^2$. Although this result is 
still preliminary, the qualitative features of the data will not change. 
Measurements on $ND_3$ have also been carried out with CLAS \cite{kuhn}, 
and on $^3He$ in JLab Hall A \cite{halla1}, to measure the 
corresponding integrals for the neutron.

\subsection{Generalized Gerasimov-Drell-Hearn sum rule for neutrons}

\begin{center}
\begin{figure}
\vspace{100mm} 
\centering{\includegraphics{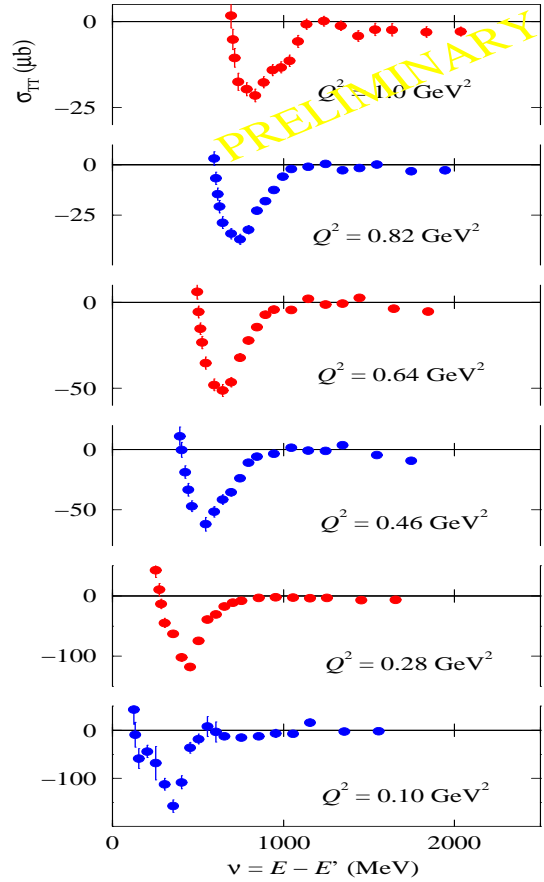}}
\caption{\small Preliminary results from experiment E94-010 on the 
integrand $\sigma_{TT}$ for the generalized GDH integral on $^3He$. 
The large negative asymmetry is due to the $\Delta(1232)$.}
\label{stthe3}
\end{figure}
\end{center}

\begin{center}
\begin{figure}
\vspace{65mm} 
\centering{\includegraphics{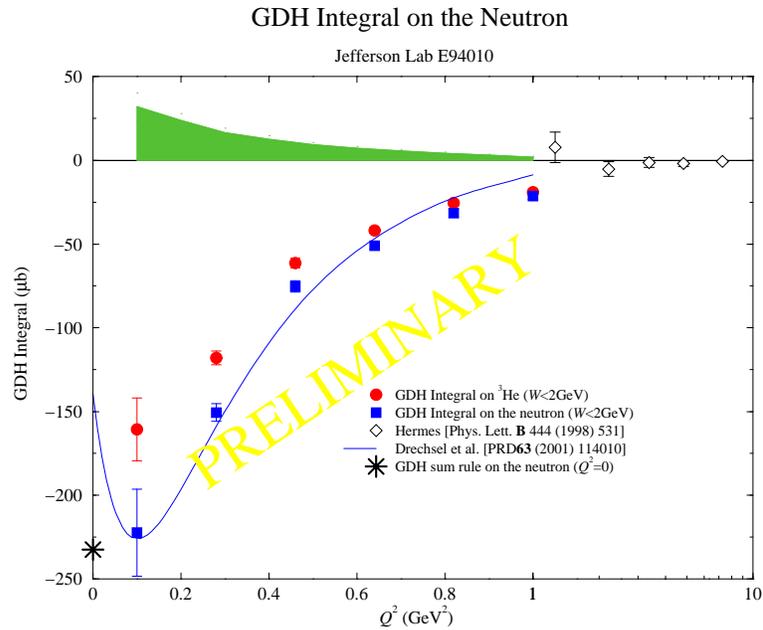}}
\caption{\small Preliminary results on the generalized GDH integral 
for $^3He$ and neutrons. The shaded 
area represents the systematic error estimate.}
\label{gdhneutron}
\end{figure}
\end{center}

Data were taken with  
the JLab Hall A spectrometers using a polarized $^3He$ target. Since the 
data were taken at fixed scattering angle, $Q^2$ and $\nu$ are correlated.
Cross sections at fixed $Q^2$ are determined by an interpolation between 
measurements at different beam energies. Both longitudinal and transverse 
settings of the target polarization were used. 
Therefore, no assumptions about $A_2$ 
are necessary in this case. 
The GDH integrand for $^3He$ is shown in Figure {\ref{stthe3}} for various 
$Q^2$. The most remarkable feature of these data is the strong negative 
contribution from the 
$\Delta(1232)$. In contrast to the proton case, the integrand above the 
$\Delta(1232)$ region remains negative and small for all $Q^2$.   
The GDH integral for $^3He$ was corrected for nuclear effects to extract the 
integral for neutrons using the prescription 
by Ciofi degli Atti \cite{ciofi}.
Preliminary results are shown in figure {\ref{gdhneutron}}. 
The integral is evaluated  
over the region from pion threshold (on a free neutron) 
to W = 2 GeV, to cover the resonance region only. 
The approach to the GDH sum rule value is slower, 
and the $Q^2$ dependence less steep than in the proton case. 
Part of this behavior is due 
to differences in the helicity structure of the dominant neutron and 
proton resonance excitations.

\section{Electroproduction of Mesons in the Nucleon Resonance Region}

A detailed study of nucleon resonance transitions requires measurement of 
exclusive final states. Current CLAS results in the region of the $\Delta(1232)$ 
and the $N^*(1535)S_{11}$ are from single $\pi^0$ and $\eta$
production, respectively. The neutral meson is inferred 
from the missing mass determined due to the overconstrained kinematics of the 
reaction. 
The search for ``missing'' resonances is systematically conducted in 
$N\pi\pi$ and $N\omega$ channels. 

\subsection{The $\gamma N\Delta(1232)$ transition multipole 
ratios $R_{EM}$ and $R_{SM}$}

The $\gamma N\Delta(1232)$ transition has been the subject of research for 
many years. 
The dominance of the magnetic dipole transition $M_{1+}$ has been known for 
three decades. The magnitudes of the quadrupole transitions, however, 
remained poorly 
determined 
until recently. The ratio  $R_{EM}=E_{1+}/M_{1+}$ was found to have 
a larger magnitude at the real photon point \cite{mami,legs} 
than constituent quark models 
predicted. New model developments that take into account explicit pion 
contributions also predict larger values, and a strong $Q^2$ dependence for the
scalar quadrupole ratio $R_{SM}=S_{1+}/M_{1+}$, while the $R_{EM}$ was 
predicted to remain nearly
constant. This made a study of the $Q^2$-dependence of the quadrupole transition 
contributions very interesting. 

\begin{figure}[t]
\vspace{90mm} 
\centering{\includegraphics{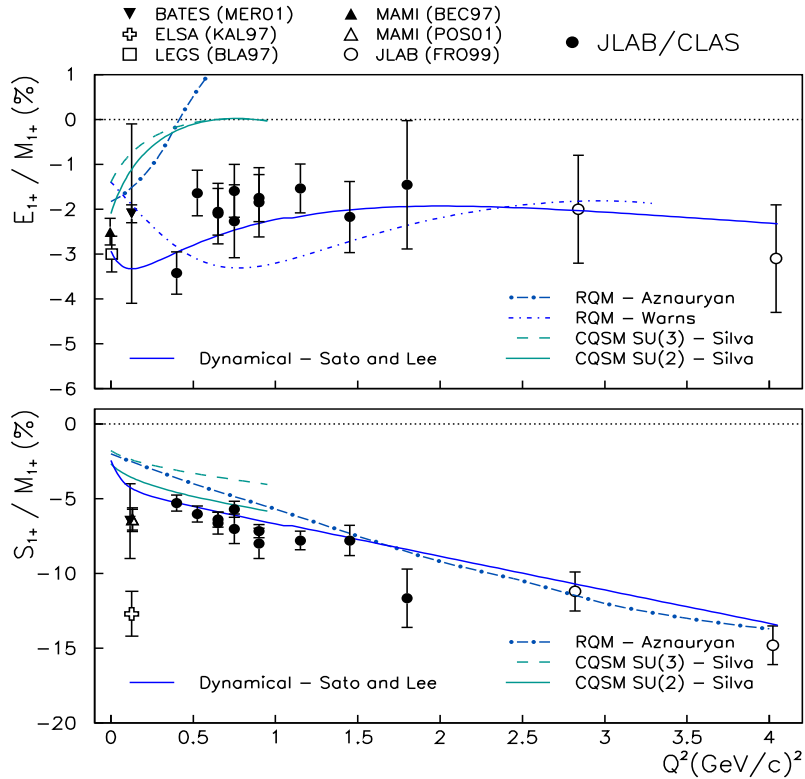}}
\caption{\small Results from CLAS on the $R_{EM}$ and $R_{SM}$ multipole 
ratios for the $\gamma N\Delta(1232)$ transition.}
\label{delta}
\end{figure}

Single pion production is most sensitive to the $\gamma N\Delta(1232)$ transition. 
The CLAS detector is well suited for this as it 
covers a large $Q^2$ and W range as well as the full azimuthal and polar 
angle distributions of the $N\pi$ system. The azimuthal distribution is fitted
to determine the response functions $\sigma_T + \epsilon \sigma_L$, 
$\sigma_{TT}$, and $\sigma_{TL}$, which are then analyzed in terms of 
multipoles. The results are
presented in Figure {\ref{delta}}. Included are various relativized quark models 
and dynamical 
models with pionic degrees of freedom. Only models that include pions 
explicitly seem to be able
to describe the $Q^2$ dependence for both the $R_{EM}$ and $R_{SM}$ simultaneously, 
while constituent quark models may describe one or the other but not both within 
the same model. It should be noted that dynamical models have been fitted to the 
photon point and to the two highest $Q^2$ data points.  
Also, chiral quark soliton models, while describing roughly the trend of 
$R_{SM}$, predict a fast falloff of $R_{EM}$ with $Q^2$ which is not 
seen in the CLAS data. 
We also do not see any trend towards significant leading order 
contributions from pQCD which require $R_{EM} \rightarrow \infty$. What is lacking 
are precise first principle QCD calculations of the $\gamma N\Delta(1232)$ 
transition multipoles.   

\subsection{The second resonance region}

\begin{figure}[t]
\vspace{90mm} 
\centering{\includegraphics{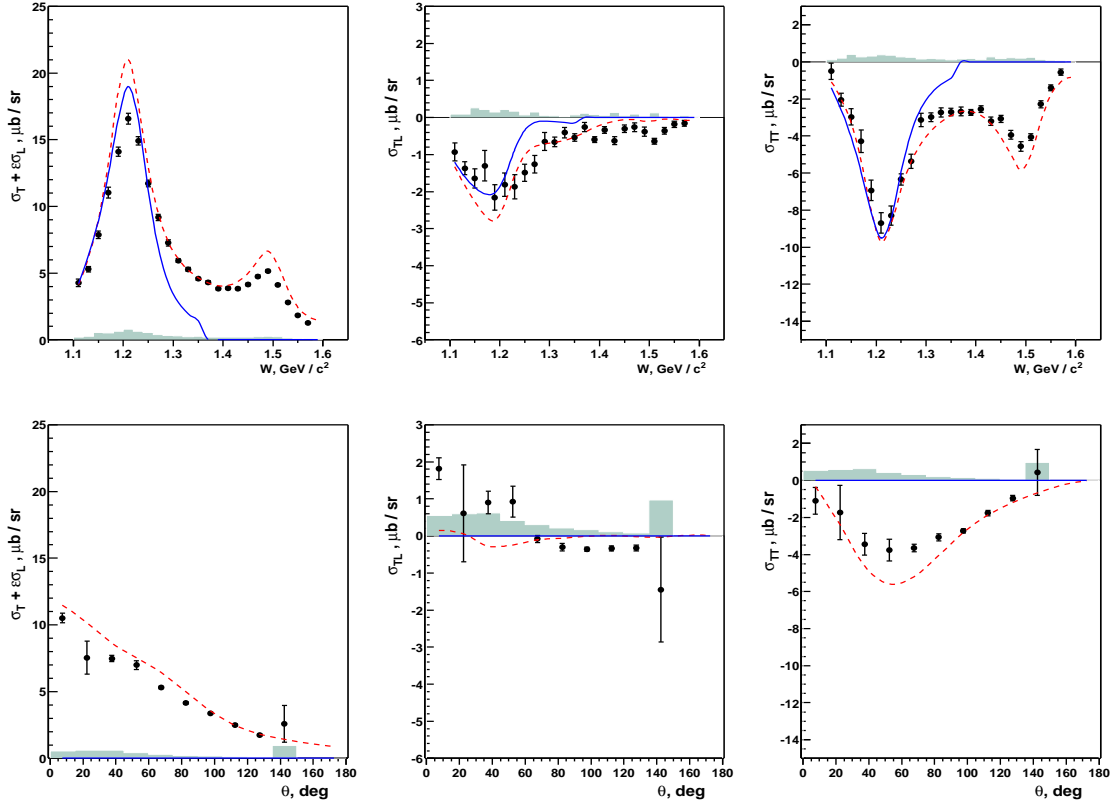}}
\caption{\small Response functions for $n\pi^+$  measured with CLAS 
at $Q^2=0.3$ GeV$^2$ and $\theta^* = 82.5^o$ as a function of W (top), 
and as a function of $\theta^*$ at W = 1.45 GeV (bottom). The dashed line 
represents the MAID2000 calculation \cite{drechsel}, the solid line is from 
the dynamical model of Sato and Lee \cite{sato}, which includes only the $\Delta(1232)$ 
as a resonant state. }
\label{npipl}
\end{figure}

A natural candidate for detailed studies beyond the $\Delta(1232)$ 
 is the Roper resonance $N(1440)P_{11}$. However, more than 35 years after
 its discovery the structure of this state is still unknown. The non-relativistic
 constituent quark model (nrCQM) puts its mass above 1600 MeV, the photocoupling 
amplitudes are not described well, and the transition form factors, 
although poorly determined, are far off. Relativized variations 
of the nrCQM improved the situation only modestly. To obtain a better 
description of the data a number of alternative models have been proposed. 
Does the Roper have a large gluonic component\cite{libuli}? Does it have a small 
quark core with a large pion cloud\cite{cano}? Or is it a nucleon-sigma 
molecule\cite{krewald}? 
It is crucial to get more precise electroproduction data, 
as it is the $Q^2$ dependence where the models differ strongly. The study of the ``Roper''
in the $p\pi^0$ channel is hampered by the presence of the dominant $\Delta(1232)$. 
Better sensitivity, due to the isospin $1\over 2$ nature of the state, should be 
obtained if the 
$n\pi^+$ channel is included in the analysis. The first $n\pi^+$ data with nearly 
complete kinematic coverage are becoming available from CLAS. Figure \ref{npipl}
shows the response functions in that channel measured throughout the first and second
resonance regions. The combined analysis of these data with the $p\pi^0$ channel 
is currently underway. 
  
Another topic in the second resonance region has been properties of the 
$S_{11}$ resonance. Analysis of single pion data gave results for the $A_{1/2}(0)$ 
photocoupling amplitude which were significantly different from what is obtained 
from the analysis of
the eta channel. More importantly, the $Q^2$ dependence of $A_{1/2}(Q^2)$ exhibits
an unusually hard transition form factor dropping by a factor of less than 
2.5 over a range $Q^2 < 3$ GeV$^2$. 
This behavior has been difficult to describe in 
quark models. In addition, the unusual $\pi N$ phase motion led to the idea
that the $S_{11}$ is not a real 3-quark resonance but possibly a $\bar{K}\Sigma$ 
molecule \cite{weise}. 
Lacking a real calculation, one might speculate that a loosely bound 
molecule would be unlikely to 
exhibit a large cross section combined with a hard transition form factor. 
Revisiting the $Q^2$ dependence of 
$A_{1/2}$ has therefore become an important topic of nucleon structure physics.

Measurements were performed with CLAS covering a range from $Q^2=0.3-4.0$ GeV$^2$.
Data below $1.5$ GeV$^2$ have been published recently \cite{thompson}. They confirm
the trend of the earlier data, showing a hard transition form factor. Preliminary new 
CLAS data covering the range $Q^2=0.2 - 3.0$ GeV$^2$ give also a very consistent 
picture \cite{dytman}, confirming the slow fall-off with $Q^2$, and linking up the photon 
data \cite{krusche} with the high $Q^2$ data \cite{armstrong}. 

There is also some good news from the theory side. The calculation by the 
Genoa group \cite{giannini} is able to reproduce the slow form factor fall-off
within a constituent quark model, using a Coulomb-type hypercentral potential and 
linear confining potential. 
The same model also describes the leading $A_{1/2}$ amplitude of the 
$N^*(1520)D_{13}$ in a large $Q^2$ range. However, the model underpredicts 
the sub-leading amplitude $A_{3/2}$. This raises the question 
whether pion cloud contributions are more prevalent in the sub-leading $A_{3/2}$
amplitudes than in the leading $A_{1/2}$ amplitudes. A dynamical model that
includes pion cloud effects could answer this question. Lattice QCD may also
be able to estimate these contributions at the photon point.    

\subsection{Missing resonances}

The so-called missing resonances \cite{isgkon} 
have been a focus of nucleon structure 
studies at intermediate energies for a number of years. It is only now that
the first experimental results have become available, and serious studies 
are being undertaken to address the issue. The importance of the topic
is due to the fact that these states are predicted within any model 
having (broken) $SU(6)\times O(3)$ symmetry, reflecting a 
symmetric arrangement 
of the 3-quark system.  Other symmetry schemes \cite{kirchbach} predict 
a smaller 
number of states, as for example a quark-diquark configuration. 
Search for at least some of the states predicted 
in one but not the other scheme is important, as it will test fundamental 
symmetry properties which are at the foundation of baryon structure in the domain
of confinement and strong QCD. Two final states, $N\omega$ and $p\pi^+\pi^-$, 
show promise in the study of higher 
mass nucleon resonances, and the search for missing states. These are currently 
under intense study with the CLAS detector.

\begin{figure}
\vspace{65mm} 
\centering{\includegraphics{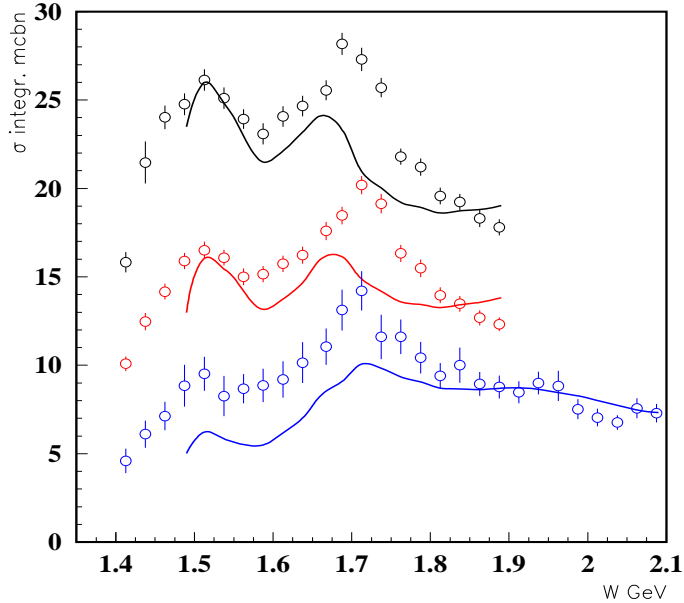}}
\caption{\small Total cross section for $\gamma^* p \rightarrow p\pi^+\pi^-$ for
different $Q^2$. The curves represent predictions based on an isobar model containing
resonance parametrizations from the analysis of single pion and eta experiments. 
The various data sets from top to bottom, correspond to 
$Q^2=0.65$ GeV$^2$, $0.95$ GeV$^2$, and $1.25$ GeV$^2$, respectively. }
\label{npipi}
\end{figure}

Figure \ref{npipi} shows total cross 
section data for the $\gamma^* p\rightarrow p\pi^+\pi^-$, showing for the first 
time resonance structure in this channel for masses greater than 1.6 GeV. The 
comparison with the model \cite{mokeev} containing the most advanced 
resonance parametrization for this mass range \cite{burk1} shows large 
missing strength in the
mass range near 1.71 GeV. While there is no missing state predicted in this mass 
range, it nevertheless shows the sensitivity of this channel to resonance
excitations. The data above 1.9 GeV are currently limited to low statistics, high
$Q^2$ data, and do not allow conclusions regarding resonance production in the 
1.9-2.1 GeV mass region where most of the missing states are predicted.     

\begin{figure}
\vspace{50mm} 
\centering{\includegraphics{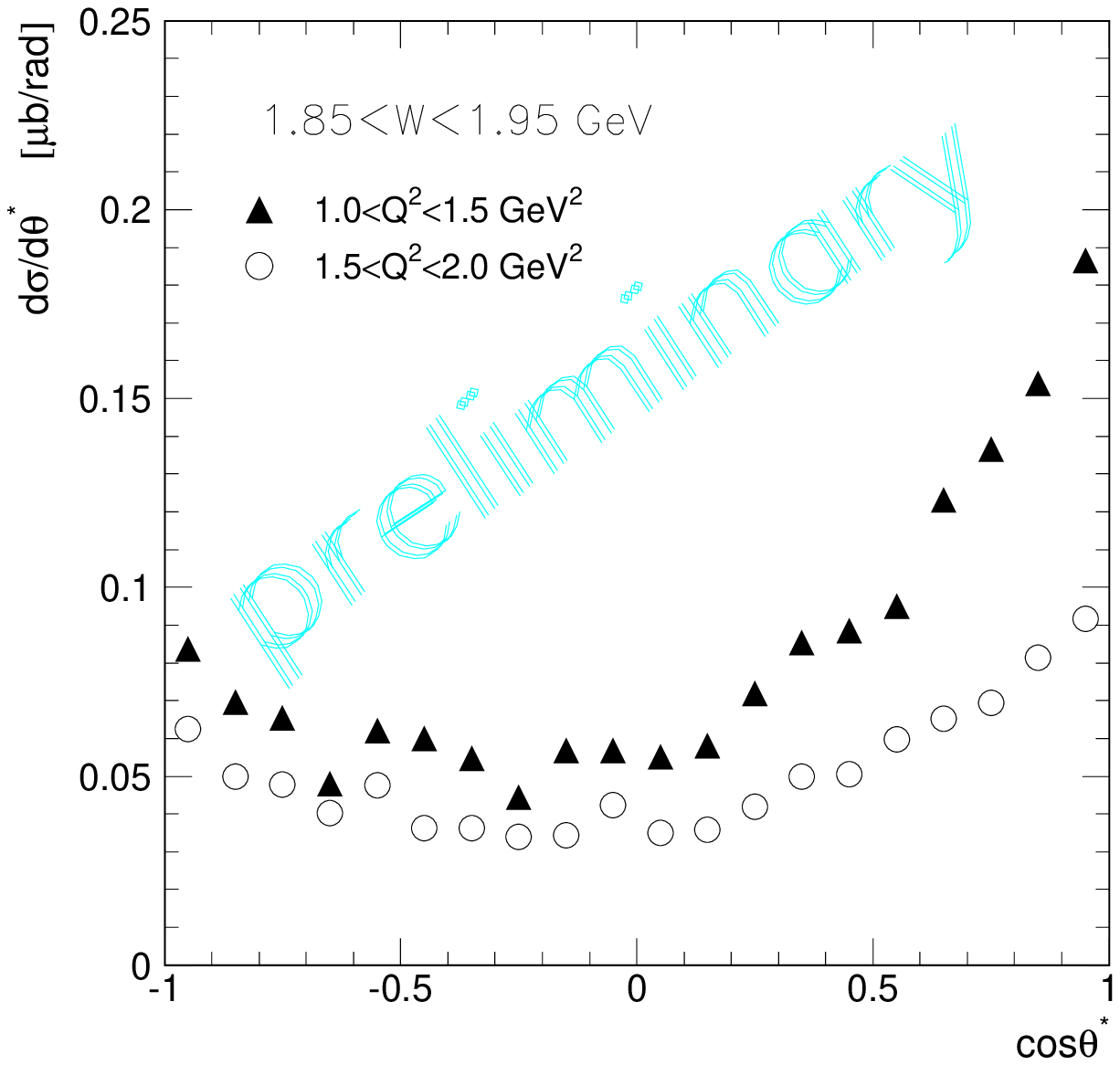}}
\caption{\small Angular distributions for $\gamma^* p \rightarrow p\omega$ at 
different values of the hadronic system W.  }
\centering{\includegraphics{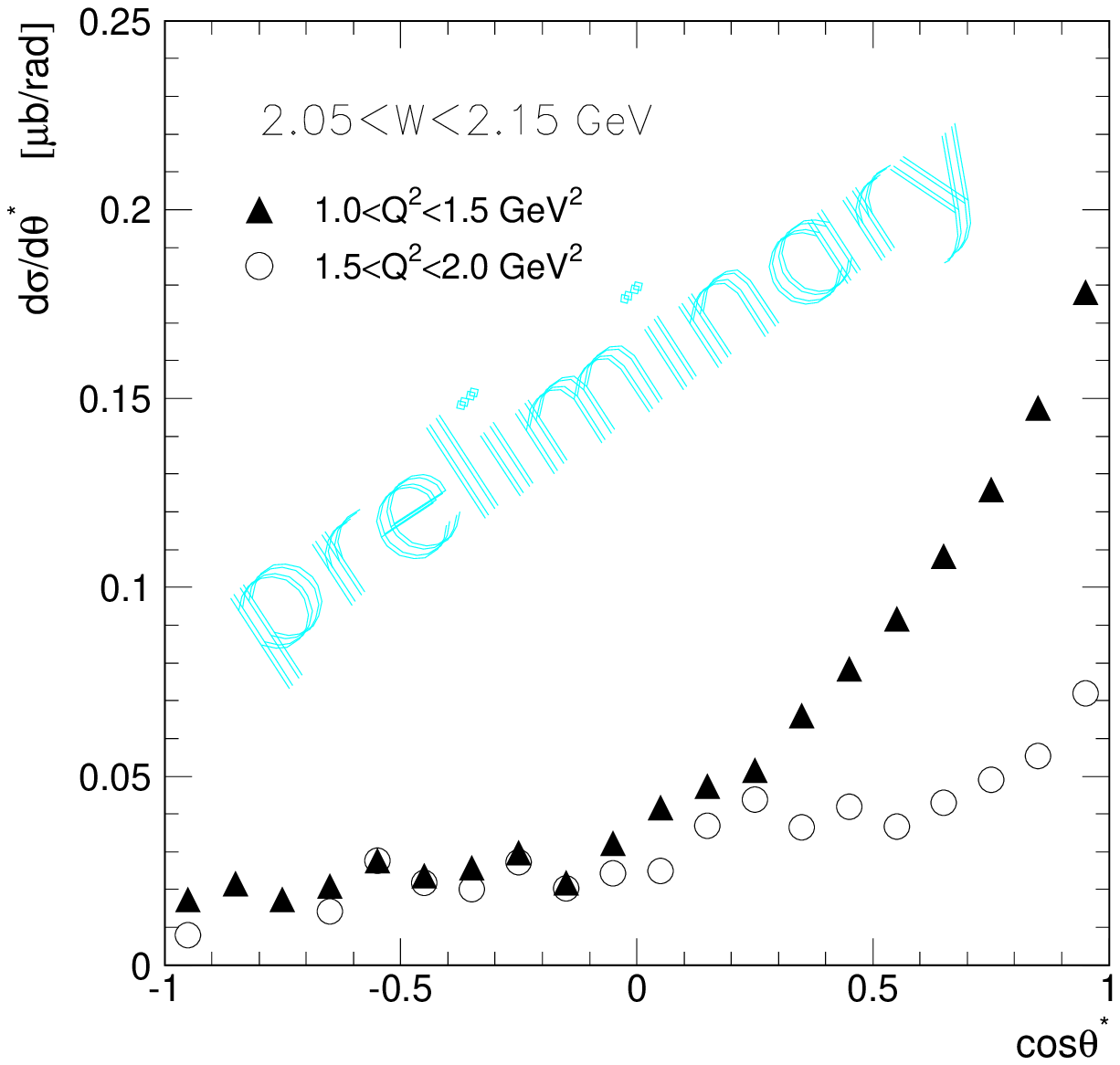}}
\label{omega}
\end{figure}

Figure \ref{omega} shows angular distributions for the $p\omega$ final state at 
different hadronic masses. This channel is expected to be dominated by t-channel
processes at forward angles and nucleon pole and resonance contributions at 
large angles. The data at high W show mostly t-channel behavior, while in the 
mass range below 2 GeV significant other contributions are visible. 
The detailed analysis of these data is currently underway. Any resonant state found
in this channel would be interesting as no nucleon resonance is currently 
known to couple to $p\omega$.

\section{DVCS - A Tool to Study Nucleon Structure}

A major goal of measuring exclusive reactions in the resonance region is 
to study the nucleon wave function which requires measurements at different 
distance scales. The interpretation 
of these reactions is complicated by the fact that the virtuality of the photon 
probe and the momentum transfer to the nucleon or excited state are strongly 
coupled leading to a correlation of the resolution of the probe and the 
momentum transfer to the recoiling baryon system. The recently established 
framework of hard exclusive reactions and generalized parton distributions 
(GPDs) offers the possiblity of studying resonance excitations where the virtuality 
of the photon probe is decoupled from the momentum transfer to the baryonic system.
For certain kinematics exclusive processes have been shown to factorize into a 
hard scattering process governed by QED and pQCD 
vertices, and the soft nucleon structure described by GPDs \cite{ji,radyushkin}.   
In the simplest reaction, the Deeply Virtual Compton Scattering (DVCS) 
$\gamma^* p \rightarrow \gamma p (\Delta , N^*)$ the virtual photon ($\gamma^*$)
has to have a sufficiently high virtuality ($Q^2$) for the process to scale.
Under these conditions the transition from the proton to the recoil baryon 
is probed at the parton level, controlled by the momentum transfer $t$, which 
can be varied independently of $Q^2$. Calculations within the GPD 
formalism for processes such as $\gamma^* p \rightarrow \gamma (\Delta(1232),
~N^*(1520),~N^*(1535))$ will be needed to enter this new area of baryon 
spectroscopy.

\begin{figure}
\vspace{70mm} 
\centering{\includegraphics{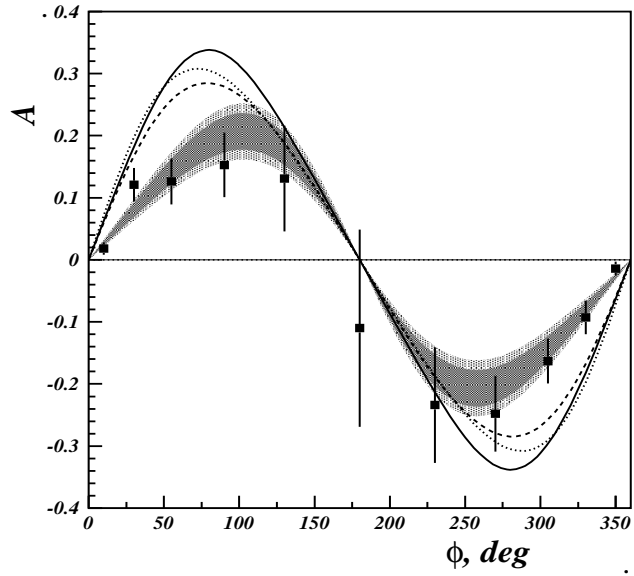}}
\caption{\small Beam spin asymmetry for the DVCS process measured with CLAS at 
4.3 GeV beam energy, $Q^2=1.3$GeV$^2$, $<t> = 0.11$GeV$^2$, $x=0.21$ . 
The curves are predictions of twist-2 and twist-3 calculations and for 
different parametrizations of GPDs.}
\label{clasdvcs}
\end{figure}

A fully exclusive measurement of the DVCS elastic process 
($\gamma^* p \rightarrow p \gamma$) was recently completed at CLAS \cite{dvcs}, using
a 4.3 GeV incident polarized electron beam. The polarized beam 
was used to exploit the interference between the DVCS and the Bethe-Heitler (BH) 
processes which results in a strong beam spin asymmetry proportional to the 
imaginary (absorptive) part of the DVCS amplitude. The results are shown
in Figure \ref{clasdvcs} in comparison with theoretical curves describing 
the reaction based on the hard scattering formalism and models for the GPDs. 

This result marks a successful foray into the uncharted territory of GPDs. 
Measurements at higher energies and with much higher statistics \cite{dvcsprop} 
are planned for the near future. Use of the inelastic DVCS process may lead to a 
promising new avenue of {\sl hard baryon spectroscopy} at the parton level.

\section{Conclusions and Outlook}

Hadron physics at JLab addresses the transition from the domain of hadronic degrees
of freedom and constituent quarks to the single parton regime. 
The first measurements of double polarization asymmetries have been 
carried out in a range of $Q^2$ not covered in previous experiments. 
The results show large contributions from resonance 
excitations with rapidly changing helicity structure. 
The first moment $\Gamma^p_1(Q^2)$ of the spin structure function 
$g_1(x,Q^2)$ shows a dramatic change with $Q^2$, 
including a sign change near $Q^2 = 0.3$ GeV. This marks the 
dominance of resonance excitations and hadronic degrees of freedom over 
the single parton domain. The $Q^2$ dependence of the generalized GDH integral for
the neutron shows dominant contributions from the $\Delta(1232)$.  In this case no
sign change is expected as the asymptotic value 
$\Gamma_1(Q^2 \rightarrow \infty) < 0$ for the neutron.

New data have been taken both on hydrogen and deuterium with nearly 10 times
more statistics, higher target polarization, and over a larger range 
of energies from 1.6 GeV to 5.75 GeV. These data will cover a  
$Q^2$ range from 0.05 to 2.5 GeV$^2$, and a larger portion of the deep 
inelastic regime. This will greatly reduce systematic uncertainties
related to the extrapolation to $x = 0$. The greatly increased 
precision, and measurements at
different energies, will give information on both $A_1$ and $A_2$.

There is also a program underway in JLab Hall A to measure the GDH 
sum rule for neutrons down to $Q^2$ values near the real photon point, 
and to measure neutron asymmetries at high $x$.    

Measurements of various exclusive processes in CLAS allows detailed studies of 
resonance excitations. Precise measurements of the transition 
multipoles in the $\Delta(1232)$ region show the importance of explicit pion 
contributions in the transition. New measurements of the $S_{11}(1535)$ transition 
form factors show a consistent behavior over the entire $Q^2$ range from 0 to
3.5 GeV$^2$.
The highly topical question of missing resonances is being addressed 
in the study of multipion and vector meson channels. Both channels show great 
sensitivity to resonance production, and structures in the data strongly 
suggest s-channel resonance contributions.

The framework of GPDs and hard scattering phenomenology has opened up a new avenue
for the study of the nucleon wave function at the parton level.

\vspace{0.5cm}\noindent
The Southeastern University Research Association (SURA) operates JLab for the 
U.S. Department of Energy under Contract No. DE-AC05-84ER40150.



\begin{thebibliography}{200} 
\bibitem{filipone}For a recent review see: B.W. Filippone, Xiangdong Ji; 
hep-ph/0101224 (2001)
\bibitem{gerasimov} S.B. Gerasimov; Sov. J. Nucl. Phys. 2, 430 (1966)  
\bibitem{drell}S.D. Drell and A.C. Hearn, Phys. Rev. Lett.16 (1966) 908
\bibitem{bjorken} J.D. Bjorken, Phys. Rev. 179, 1547 (1969)
\bibitem{jios} X. Ji, J. Osborne, J. Phys. G27:127 (2001) 
\bibitem{abe} K. Abe et al., Phys. Rev. D58, 2003 (1998)
\bibitem{burli} V. Burkert and Zh. Li, Phys. Rev. D47,46 (1993)
\bibitem{soffer} J. Soffer and O.V. Teryaev, Phys. Rev. Lett. 70, 3371 (1993)
\bibitem{buriof} V. Burkert and B. Ioffe, Phys. Letts. B296, 223 (1992); J.Exp.Theo.Phys.78, 619 (1994) 
\bibitem{xji} X. Ji, C.W. Kao, J. Osborne, Phys. Lett. B472:1-4 (2000) 
\bibitem{burk} V. Burkert, Phys. Rev. D63, 97904(2001)  
\bibitem{burk1} V. D. Burkert, Czech. Journal of Physics,Vol. 46, 627, 1996.
\bibitem{bucramin} V. Burkert, D. Crabb, R. Minehart, et al., JLab experiment 91-023.
\bibitem{kuhn} S. Kuhn, G. Dodge, M. Taiuti, et al., JLab experiment 93-009
\bibitem{halla1} Z.E. Meziani et al., JLab experiment E94-010
\bibitem{ciofi}C. Ciofi degli Atti, S. Scopetta, Phys. Lett. B404, 223-229 (1997)   
\bibitem{mami} R. Beck et al., Phys. Rev. Lett. 78:606-609 (1997); Phys. Rev. C61:035204 (2000)
\bibitem{legs}G. Blanpied et al., Phys. Rev. C64:025203 (2001)   
\bibitem{drechsel} D. Drechsel, O. Hanstein, S. Kamalov, L. Tiator, Nucl. Phys. A645, 145 (1999) 
\bibitem{sato} T. Sato, T.-S.H. Lee, Phys. Rev. C63, 055201 (2001) 
\bibitem{libuli} Zp. Li, V. Burkert, Zh. Li, Phys. Rev. D47, 46 (1993) 
\bibitem{cano} F. Cano and P. Gonzalez, Phys. Lett. B431, 270-276 (1998)
\bibitem{krewald} O. Krehl, C. Hanhart, S. Krewald, J. Speth, Phys. Rev. C62, 025207 (2000)   
\bibitem{weise} N. Kaiser, T. Waas, and W. Weise, Nucl. Phys. A612, 297 (1997) 
\bibitem{thompson} R. Thompson et al. (CLAS coll.), Phys. Rev. Lett. 86:1702-1706 (2001)  
\bibitem{krusche} B. Krusche et al., Phys. Rev. Lett. 74, 3736 (1995)
\bibitem{armstrong} C. S. Armstrong et al., Phys. Rev. D 60, 052004 (1999)
\bibitem{dytman} S. Dytman, private communications
\bibitem{giannini} M.M. Giannini, E. Santopinto, Few Body Syst. Suppl. 11:37-42 (1999) 
\bibitem{isgkon} N. Isgur and R. Koniuk, Phys. Rev. Lett. 44, 845-848 (1980)
\bibitem{kirchbach} M. Kirchbach, Mod. Phys. Lett. A12:3177-3188 (1997) 
\bibitem{ao} V.D. Burkert, Zh. Li, Phys. Rev. D46,47 (1993); R. DeVita, private 
communication.
\bibitem{mokeev} M. Ripani, et al., Nucl.Phys.A672:220-248,2000 
\bibitem{ji} X. Ji, Phys. Rev. Lett. 78, 610 (1997); Phys. Rev. D55, 7114 (1997)
\bibitem{radyushkin} A. Radyushkin, Phys. Lett. B380, 417 (1996), 
Phys. Rev. D56, 5524 (1997)
\bibitem{dvcs} S. Stepanyan, V. Burkert, L. Elouadrhiri,  et al., hep-ex/0107043
\bibitem{dvcsprop} V. Burkert, L. Elouadrhiri, M. Garcon, S. Stepanyan, et al., 
JLab experiment E01-113.
\end{thebibliography}
\end{document}